This is a file written in LaTex Version 2.09. There is one figure available
upon request from the authors on the email address: vagner@eta.pha.jhu.edu .

\documentstyle[titlepage,12pt]{article}
\begin{document}
\parindent 2em
\baselineskip 4.5ex

\begin{titlepage}
\begin{center}
\vspace{12mm}
{\LARGE Density-functional theory of freezing of vortex-liquid in quasi
two-dimensional superconductors}
\vspace{25mm}

Igor F. Herbut and Zlatko Te\v sanovi\' c \\
Department of Physics and Astronomy, Johns Hopkins University, Baltimore,
MD 21218

\end{center}
\vspace{10mm}

\noindent
{\bf Abstract:}
We present a theory of vortex liquid-to-solid
transition in homogeneous quasi
2D superconductors.
The free energy is written as a functional of density of zeroes
of the fluctuating order parameter.
The transition is weakly
first-order and well below the $H_{c2}(T)$ line.  Transition
temperature, discontinuities of the average
Abrikosov ratio and of the average superfluid density, the Debye-Waller
factor and the latent heat are
in good agreement with
Monte Carlo simulations.  The density is only weakly modulated
in the ``vortex-solid" phase, consistent with the
density-wave behavior.


\end{titlepage}

The experimental observation\cite{1} of the wide region around the $H_{c2} (T)$
line where the Abrikosov
lattice appears melted
demonstrates the importance of thermal fluctuations in high
temperature superconductors (HTS).
Such fluctuations are generally important in quasi
two-dimensional (2D) materials, such as thin films or layered systems
with weak interlayer coupling.  The
existence and the nature of ordered low-temperature phase
is still matter of some controversy.
The early theories \cite{2} {\it assumed} the
ordered phase  at low temperatures and proposed dislocation-unbinding
as a mechanism of second-order melting transition
or used the Lindemman criterion to locate the melting point \cite{3}. However,
even the very existence of vortex-lattice phase, which is the starting point
of these theories, cannot be taken for granted \cite{4,5}. External magnetic
field causes the system to appear D-2 dimensional on perturbative level,
and the  perturbation theory carried out to eleventh order \cite{6}
shows no sign of low-temperature Abrikosov phase in 2D. Renormalization group
analysis near the upper critical dimension D=6 found the continuum of
relevant charges to be generated by the renormalization procedure and
indicated that the transition into Abrikosov vortex-lattice state might be
first order, as expected from the simple
Landau argument\cite{7}. The relevance of this analysis for
real superconductors,
however, is not obvious
since the lower critical dimension
for the Abrikosov transition is D=4.
Recently, Monte Carlo (MC)
simulations \cite{8,9,10} revealed very weak
first order vortex liquid to solid
transition in 2D.  At present, our information about this transition
comes almost exclusively from such numerical work.

The main purpose of this paper is to develop a
working {\it analytic} theory which has predictive
power and which can be applied to a wide variety of
experimentally relevant issues.
We derive a density-functional representation of the free energy for
zeroes of the fluctuating superconducting order
parameter, with the exact two-body correlations built in.
Our theory is a ``first principle" in the sense that it can
be obtained from the microscopic description by a
series of well understood approximations.
It is then shown that above certain values of
the first two peaks in the vortex
structure factor triangular lattice has lower free energy than the uniform
configuration. Using the information on the
dependence of the structure factor on
temperature from the existing MC simulations we locate the
transition temperature and calculate several typical quantities like
the latent heat and the change in
thermal average of the Abrikosov's ratio at the transition. Obtained results
are in good agreement with those
found in numerical simulations.

The starting point is the Ginzburg-Landau (GL)
partition function for a homogeneous 2D superconductor in perpendicular
magnetic field, with fluctuations of the magnetic field neglected
($\kappa \gg 1$). We are interested in  the regime where the Landau level (LL)
structure of Cooper pairs dominates the fluctuation spectrum:  This is the case
for
fields above $H_b\approx (\theta/16)H_{c2}(0)(T/T_{c0})$,
where $\theta$ is the Ginzburg fluctuation parameter \cite{11} (For
example, in BSCCO 2:2:1:2, $\theta\sim 0.045$ and $H_b\sim 1$ Tesla).
In this regime, the
approximation in which one retains only the lowest Landau level
(LLL) modes describes essential features of the problem.
The partition function is then $Z=\int D[\psi^{*} \psi] \exp{(-S)}$, and
\begin{equation}
S=
\frac{d}{T}\{ \alpha '(T) \int d^{2}\vec{r}
|\psi(\vec{r})|^{2} + \frac{\beta}{2}
\int d^{2}\vec{r} |\psi(\vec{r})|^{4} \}~~~,
\end{equation}
where $\alpha '(T) =\alpha(T)(1-H/H_{c2}(T)) $, $d$ is the thickness of the
film
 or the effective interlayer separation
and $\alpha(T)$ and $\beta$ are phenomenological parameters.
We choose to work in the
symmetric gauge so that the order parameter is a holomorphic function
and can be written as
$\psi(z)=\phi
\prod_{i=1}^{N} (z-z_{i})\exp{-(|z|^{2}/4)}$
where $N$ is the area of the system in
units of $2\pi l^{2}$, $z=(x+iy)/l$ and $l$ is the magnetic length for
the charge $2e$.  The partition function can now be expressed in terms
of variables $\phi$ and \{$z_{i}$\} \cite{8}. They
represent two distinct tendencies in a
superconductor; $\phi$ describes the overall growth of the local
superconducting
order, while \{$z_{i}$\} represent the remaining weak lateral
correlations between vortices.  If the latter are treated in a certain
average way one can
account for 98\% of the
thermodynamics \cite{11}. However,
the transition that we want to discuss is entirely within the remaining 2\%,
and the correlations among $\{ z_i\}$
are crucial.
The integration over $\phi$ can be performed exactly in the thermodynamic
limit $N\rightarrow \infty$, yielding
$Z=\int \prod_{i} (dz_{i}dz_{i}^{*}/2\pi) \bar{f^{4}}^{-N/2} \prod
|z_{i}-z_{j}|^{2} \exp\{-S'\}$, where
\begin{equation}
S'=-\frac{V^{2}}{2}+\frac{V}{2}(V^{2}+2)^{1/2} + \sinh^{-1}(V/\sqrt{2})),
\label{2}
\end{equation}
$V(\{z_{i}\})=g\bar{f^{2}}/\sqrt{\bar{f^{4}}}$, $\bar{f^{2}}(\{z_{i}\})
=\int (dz dz^{*}/2\pi N) \prod_{i} |z-z_{i}|^{n} \exp {-(n|z|^{2}/4)}$
and $g=\alpha ' \sqrt{\pi l^{2}d/
\beta T}$. The variable $\phi$ is determined by
\begin{equation}
<|\phi|^{2}> \bar{f^{2}}\frac{2\pi l^{2}d|\alpha'|}{T}=
V^{2}(1+\sqrt{1+2V^{-2}}).
\label{3}
\end{equation}
The original problem is now equivalent to the thermodynamics
of classical 2D system of particles which we call dense-vortex-plasma (DVP)
\cite{8}.
We note several features of DVP system which should be reflected
in our density-functional:
1) particles
interact via long-range multiply-body forces, 2) the system
is incompressible, 3) the system is scale invariant,
4) the thermodynamics
depends on a single dimensionless coupling constant $g$. Previous MC
simulations show  transition from a liquid  to a solid state taking
place
at $g_{F}\approx -7$ \cite{8,9,10}. The mechanism of the transition,
however, remains
obscure in the partition function (2) since the exponentiated energy is still a
perfectly smooth function of $V(\{z_{i}\})$.
 From Eq. (2) we can write the {\it exact} free energy per vortex as
$f(g)/T = S'-s(<\beta_{A}>)$
where $<\beta_{A}>$ is the $g$-dependent
thermal average of the Abrikosov ratio
$\beta_{A}(\{z_{i}\})=\bar{f^{4}}/(\bar{f^{2}})^{2}$ determined
by minimizing $f(g)$. The entropy $s(<\beta_{A}>)$ is given by
\begin{equation}
s(<\beta_{A}>)=N^{-1}\ln \int \prod_{i} \frac{dz_{i}dz_{i}^{*}}{2\pi l^{2}}
\bar{f^{4}}^{-N/2} \prod_{i<j} |z_{i}-
z_{j}|^{2} \delta(\frac{\bar{f^{4}}}{(\bar{f^{2}})^{2}}-<\beta_{A}>).
\label{4}
\end{equation}
The transition is (as usual) hidden in the entropy part of the free energy,
but the evaluation of the integral (4) is a hopeless task. Obviously, a
different
route needs to be taken.

To study solidification transition it is often beneficial to change
variables from particle coordinates to density of particles. This can be
achieved by inserting the unity
$1=\int D\rho(\vec{r})  \delta[\rho(\vec{r})-\sum \delta(
\vec{r}-\vec{r}_{i})]$ in $Z$ (2). After the $\delta$-function
is expressed as an integral over auxiliary field $\Phi(\vec{r})$ coordinates
of vortices can be integrated out.
The partition
function becomes $Z=\int D\rho(\vec{r}) D\Phi(\vec{r}) \exp \{-S''\}$ ,
\begin{equation}
S''=U[\rho(\vec{r})]-\int d^{2}\vec{r} \Phi(\vec{r}) \rho(\vec{r})
-N \ln \int d^{2}\vec{r}\exp{(-\Phi(\vec{r}))}
\label{5}
\end{equation}
where $U[\rho]$ is the energy density-functional determined by the details
of the DVP interaction and we changed variables $i\Phi\rightarrow\Phi$.
 We expand $U[\rho]$ in terms of multy-body interactions
and keep only the first,
 two-body term  so that $U=\frac{1}{2}\int \rho(\vec{r_{1}})
 v(\vec{r_{1}}-\vec{r_{2}})\rho(\vec{r_{2}}) + O(\rho
^{3})$. After this truncation of the energy functional,
density can be integrated out leaving the
partition function as an integral over field $\Phi(\vec{r})$ with the action
\begin{equation}
F=-\frac{1}{2}\int d\vec{r_{1}}d\vec{r_{2}}
\Phi(\vec{r_{1}}) v^{-1}(\vec{r_{1}}-\vec{r_{2}})
\Phi(\vec{r_{2}}) - N \ln \int d^{2}\vec{r} \exp{(-\Phi(\vec{r}))}.
\label{6}
\end{equation}
{\it We take the Eq. (6)
to be the desired expression for the  mean-field
free energy of the vortex system}. In order to account
for the strong two-particle correlations that exist in the liquid state
of DVP we make the
 substitution $-v(r)=c(r)$,
where $c(r)$ is the direct (Ornstein-Zernike)
correlation function of the vortex-liquid whose Fourier transform is
related to the structure factor via the relation
$S(q)=(1-c(q))^{-1}$ \cite{12}. The
structure factor $S(q)$ is the Fourier transform of the connected
density-density correlation function.

Before proceeding with the analysis of the phases described by the free
energy (6) let us comment on validity of the truncation of the
DVP energy density functional. Even if the particles interacted only via
two-body forces all multi-body correlations will be generated and the
true free energy would contain additional terms in (6) of higher order in
field $\Phi(\vec{x})$. Therefore,
the fact that particles in DVP interact through
multi-body forces makes the approximate free energy (6) no less appropriate
here than for the ordinary correlated liquid.  In fact one might expect
that the long-range interaction operating in DVP will make the system only
more mean-field like. The important feature of the approximate free energy (6)
is the last term which is purely entropic and
hence crucial for the freezing transition. The free energy (6) may be thought
of as a harmonic approximation around the liquid state and it will be shown
that this particular form of it is very  good for an incompressible
system.

We assume that the freezing of the DVP happens into triangular lattice
with period $a$, $a^{2}=4\pi/\sqrt{3}$ and we work in
units  $l=1$.  Note that this assumption is not necessary and
that we could consider other lattices.  Actually, at this
point it is not obvious that the ordering wave-vector
will be in any simple relationship to the magnetic length.  Within
the resolution of the MC data for the structure factor,
however, the strongest tendency
for ordering corresponds to the triangular modulation.
We determine the solid phase by two real order
parameters $\Phi_{1}$ and $\Phi_{2}$ in the following way:
$\Phi(\vec{r})=\Phi_{0}-2\sum_{i=1}^{2}\Phi_{i} f_{i}(\vec{r})$,
where $f_{1}(\vec{r})=1/2\sum_{G_{1}}
\exp{i\vec{G_{1}}\vec{r}}=\cos(ax)+2\cos(a\sqrt{3}y/2)\cos(ax/2)$
and the sum runs over six shortest reciprocal lattice vectors $|\vec{G}|=
G_{1}=a$. The same sum,
but over next six shortest reciprocal vectors  $|\vec{G}|=G_{2}=\sqrt{3}a$
determines  the function $f_{2}(\vec{r})=\cos(a\sqrt{3}
y)+2\cos(3ax/2)\cos(a\sqrt{3}y/2)$. Inserting this into expression (6) we get
two-order-parameter form of the free energy per vortex
\begin{equation}
F=\frac{\Phi_{0}^{2}}{2c_{0}}+
\Phi_{0}+3\sum_{i=1}^{2}\frac{\Phi_{i}^{2}}{c_{i}}
- \ln \int d^{2}\vec{r} \exp{(2\sum_{i=1}^{2}\Phi_{i}f_{i}(\vec{r}))}
\label{7}
\end{equation}
 where $c_{1/2}=c(G_{1/2})$. The
 free energy is minimized by the solutions of the
equations:
\begin{equation}
\frac{3\Phi_{j}}{c_{j}}=\frac{\int d^{2}\vec{r}f_{j}(\vec{r}) \exp{(2\sum_{i=1}
^{2}\Phi_{i}f_{i}(\vec{r}))}}
{\int d^{2}\vec{r}\exp{(2\sum_{i=1}^{2}\Phi_{i}f_{i}(\vec{r}))} }
\label{8}
\end{equation}
for $j=1,2$. The equation for the uniform component is $\Phi_{0}=-c_{0}$  and
system is manifestly incompressible since the average
density $\rho(\vec{G}=0)=1$ does not change with temperature.
 It is easily seen that $\int f_{j}(\vec{r})=0$
 and one solution of the equations is
always $\Phi_{1}=\Phi_{2}=0$ which
describes the liquid state of DVP. To find a non-zero solution, we choose
a pair $(c_{1},c_{2})$, numerically
solve Eqs. (8) and compare free energies of
the solid and the liquid configurations.
 This way one obtains the crosses joined joined by the full line with a
negative slope on the structural phase diagram
presented on Fig. 1. For values of $S_{1}$ and $S_{2}$ above the
line triangular lattice of vortices is the stable phase, while below it
DVP is in liquid phase.

To find the value of the coupling constant $g_{F}$ where the DVP freezes
the information on $S_{1}$ and $S_{2}$ as functions of $g$ is needed.
The vortex structure factor can be found by using
the perturbation theory to determine the structure factor for
the superfluid density, $|\psi (\vec r)|^2$, and then connecting
the two through the expression for $\psi (\vec r)$ given below Eq. (1)
and some of the standard correlation function hierarchies (BBGKY,
hypernetted chain, etc.).  While this is possible in principle it
is more expedient here to utilize the vortex structure factors from
MC simulations \cite{8,13}.  Particularly useful is
a detailed analysis of the peaks in the
vortex structure factor
by O'Neill and Moore \cite{13}. In the MC simulations in the spherical
geometry they find that, for $5<g^{2}<50$, both peaks depend
approximately linearly on
$g^{2}$. The straight line with positive slope in Fig. 1 represents
$S_{1}-S_{2}$ relation derived from there. Two lines intersect at $S^{F}_{1}=
4.45$ and $S^{F}_{2}=1.47$, which gives $g_{F}=-6.5$ and the values
of order parameters at the transition are $\Phi_{1}=0.50$ and $\Phi_{2}=0.10$.
Interestingly,
around this value of the coupling constant
O'Neill and Moore first start to see critical slowing down of the dynamics
in their MC simulations,
but they argued against finite temperature phase transition. Our result
is in excellent agreement with MC simulations in Refs. \cite{8,9,10}.

The main features of DVP (long-range
interactions, incompressibility) resemble those
of another well studied classical system, namely 2D one component Coulomb
plasma
\cite{14}. It is therefore not surprising that the values of first
two peaks of the structure factor that we found
at the transition closely match the numbers found in MC simulations there.
In particular, the value of $S_{1}^{F}$ is very close to
4.4 which is known as 2D version of the Verlet criterion. For comparison,
our number for $S^{F}_{2}$ is much worse for the hard-disk system  where it
equals 1.9 \cite{14}. This is
because the system of hard-disks is compressible and the proposed form of the
mean-field
free energy (6) ignores the
change in average density at the freezing transition.

Since the transition takes place at a finite value of $S_{1}$ the transition
is first order in agreement with what has been observed in
recent experiments in clean YBaCuO single
crystals
\cite{15}. The latent heat comes solely from the structural change at the
transition. If we take the expression for entropy of the system to be
$ s=\int d^{2}\vec{r} \rho(\vec{r}) ln [\rho(\vec{r})] $ the latent heat
equals $Q=NT_{F}3(\Phi_{1}^{2}/c^{F}_{1} + \Phi_{2}^{2}/c^{F}_{2})$. For the
above numbers this gives
$Q/NT_{F}=
1.0$, somewhat higher than the number 0.4 found in MC simulations
\cite{9,10,14}. Going back to the exact expression for
the free energy it follows that the thermal average of
Abrikosov ratio has a discontinuity at the transition
$\Delta <\beta_{A}>/<\beta_{A}>\approx<\beta_{A}>
\Delta Q /g_{F}^{2}NT_{F}=0.025$ if we
take $<\beta_{A}>\approx1.2$ at $g=g_{F}$.
Invoking Eq. (3) we see that the average superfluid density has a
discontinuous jump of about $2\%$. This
in reasonable agreement with the results of Refs. \cite{9} and \cite{10}.
The Fourier components of the density
of vortices can be determined from :
\begin{equation}
\rho(\vec{G}) = \frac{\int d^{2}\vec{r} \exp(i\vec{G}\vec{r}-\Phi(\vec{r}))}
{\int d^{2}\vec{r} \exp(-\Phi(\vec{r}))}
\end{equation}
and the Debye-Waller factor is given by
$\nu(G)=|\rho(G)|^{2}$. We find that for
$|\vec{G}|\ge G_{1}$ $\rho(G)\approx 0.72
\exp(-\lambda ^{2} G^{2})$ with $\lambda=0.47$.
Similar Gaussian fall-off was found for freezing into 3d bcc or fcc lattices,
but with smaller coefficient $\lambda$
($\lambda_{bcc}=0.34, \lambda_{fcc}=0.19$, \cite{12}).
The fast decay of Debye-Waller factor indicates that the transition is very
weakly first order and the density modulation is  rather weak right below the
transition.

The present scheme allows us to examine the possibility of
high temperature solid phase.
We determined the values of $(c_{1},c_{2})$  where the local $\Phi_{1}\ne 0$,
$\Phi_{2}\ne 0$ minimum of the free energy first appears; at that point
free energy of the liquid state is still lower than the one of the solid.
 This way triangles joined by the
dashed line on Fig. 1 are obtained. Intersection with the $S_{1}-S_{2}$ line
then determines the highest value of $g$ where the solid is still
locally stable:
$g_{sh}=-6.25$. The interval $\Delta g_{sh}=g_{sh}-g_{F}=0.25$ agrees well with
 the value 0.2 obtained in Ref. \cite{9} for the finite size system.
The supercooling
of the liquid phase is also possible, and the lowest $g=g_{sc}$
 where the liquid
phase is locally stable is given by the value where the first peak of
$S(\vec{q}
)$ diverges. From the data in Ref. \cite{9} we expect
that the interval $\Delta g_{sc}=g_{F}-g_{sc}$
 is of the same size as $\Delta g_{sh}$. The supercooling and the
superheating data from the same reference indicate that there is a well defined
critical value of Abrikosov ratio $<\beta_{A}>\approx 1.18$ at which
the DVP system can undergo a continuous transition.

The vortex-liquid freezing transition discussed here is not a
superconducting transition in the sense of breaking of U(1)
symmetry \cite{5}. It is best thought of as a
transition to a charge-density-wave of Cooper pairs. This is
consistent with the weak density modulation of the ordered
phase.  The superconducting pairing susceptibility remains
short-ranged in this density-wave, although this range becomes
$\gg l$ as density modulation increases.  This phase
should be distinguished from other phases described
in the literature \cite{16}. True superconducting transition in
high magnetic field is almost always due to
disorder, and, unless the disorder is very strong, it will be much
below the freezing transition discussed here.  Finally, the
weak interlayer coupling in HTS and similar systems will act
to ``lock-in" density-waves from different layers resulting
in a 3D-ordered weakly modulated structure.  The first
order character of the transition will be enhanced.  The change in the
transition temperature will be rather small and can be
easily estimated \cite{8}.

To conclude, we have presented a density-functional
theory of
the vortex liquid-to-solid phase transition
in homogeneous quasi 2D
superconductors.
As the input, we have used the temperature-dependent first two peaks of the
numerically determined structure factor for
the zeroes of the superconducting order parameter.
As the output, we obtain the value of transition temperature,
the discontinuities of thermal average of
the Abrikosov ratio and of the average superfluid density,
the Debye-Waller factor at
the point of transition and the latent heat. We determine the temperature
interval for superheating of the vortex-solid phase and conjectured that
there is a critical point that can be reached by supercooling.
The results are found to agree well
with Monte Carlo simulations.
While we have used the numerical results as an input to
our theory, it is not necessary to do so.  Large order
perturbation expansion results for the structure factor of
$|\Psi (\vec r)|^2$ \cite{17} could have been used instead so
that the theory is kept entirely at the analytic level.

This work has been supported in part by the David and Lucile Packard
Foundation.
One of us (IFH) thanks Professor
M. O. Robbins for useful discussions in the early stage of this work.

\pagebreak
Captions:

Figure 1. Structural phase diagram for vortex system in terms of first two
peaks
of the structure factor of vortices.
Crosses denote the values of $S_{1}$ and $S_{2}$
where the vortex-solid becomes energetically favorable, and triangles
mark the points where the solid first becomes locally stable.  The line with
positive slope depicts $S_{1}-S_{2}$ relation inferred from MC simulations
\cite{13}. The transition occurs at the intersection of this line with the
line which joins the crosses.

\pagebreak

\end{document}